 \documentclass [12pt,a4paper,     ]{article}
\usepackage{times}

\DeclareFontFamily{OT1}{times}{}
\DeclareFontShape {OT1}{times}{m }{n }{ <-> ptmr }{}
\DeclareFontShape {OT1}{times}{bx}{n }{ <-> ptmb }{}
\DeclareFontShape {OT1}{times}{m }{it}{ <-> ptmri}{}
\DeclareFontShape {OT1}{times}{bx}{it}{ <-> ptmbi}{}
\usepackage{amsmath}
\usepackage{amsfonts}
\usepackage{amssymb}
\usepackage{latexsym}

\newcommand{\cl}{C \kern -0.1em \ell_\pi} % Clifford algebra
\newcommand{\CON}{\overline}          % quaternion (vector)conjugate

\newcommand{\Scal}{\mathbb{S}}        % scalar part unary  S[AB] = s
\newcommand{\vect}{\wedge}            % vector part binary   A^B = v

\newcommand{\VEC}{\vec{\kern +.1em[}} % VEC-tor left  bracket   |>.. 
\newcommand{\TOR}{\vec{\kern +.2em]}} % vec-TOR right bracket   ..>| 
\newcommand{\BRA}{\langle\kern -.2em\langle} % Dirac BRA      <<...| 
\newcommand{\KET}{\rangle\kern -.2em\rangle} % Dirac KET      |...>> 

\newcommand{\REA}{\operatorname{Re}}  % real      part
\newcommand{\Oh}{\tfrac{1}{2}}        % 1/2 = 0.5
\newcommand{\Q}{[\hspace{1.mm}]} % quaternion operand A{\Q}B -> A[]B
\begin{document}

\noindent {\bf Published in \emph{Foundations of Physics},
                             Vol. 35, No. 5, May 2005, 865--880.}

\vspace{8\baselineskip}

\noindent {\bf {\LARGE \medskip Cornelius Lanczos's derivation of the usual\\ action integral of classical electrodynamics}}

\vspace{1\baselineskip}

{\bf \noindent Andre Gsponer and Jean-Pierre Hurni}\footnote{Independent Scientific Research Institute, Box 30, CH-1211 Geneva-12, Switzerland; e-mail: isri@vtx.ch}

\vspace{3\baselineskip}

{\small \noindent \emph{The usual action integral of classical electrodynamics is derived starting from Lanczos's electrodynamics --- a pure field theory in which charged particles are identified with singularities of the homogeneous Maxwell's equations interpreted as a generalization of the Cauchy-Riemann regularity conditions from complex to biquaternion functions of four complex variables.   It is shown that contrary to the usual theory based on the inhomogeneous Maxwell's equations,  in which charged particles are identified with the sources, there is no divergence in the self-interaction so that the mass is finite, and that the only approximations made in the derivation are the usual conditions required for the internal consistency of classical electrodynamics. Moreover, it is found that the radius of the boundary surface enclosing a singularity interpreted as an electron is on the same order as that of the hypothetical ``bag'' confining the quarks in a hadron, so that Lanczos's  electrodynamics is engaging the reconsideration of many fundamental concepts related to the nature of elementary particles.}}

\vspace{2\baselineskip}


\vspace{1\baselineskip}

\vspace{2\baselineskip}
{\bf \noindent 1. INTRODUCTION}
%------------------------------
\vspace{1\baselineskip}
\label{int:0}

One of the most fundamental equations of present-day physics, from classical electrodynamics to the Standard model of elementary particles, is the action integral for a charged particle in an external electromagnetic field
\begin{equation} \label{int:1}
   S = \int_{\tau_1}^{\tau_2} d\tau~ \Bigl[
                 - mc^2
                 - e \gamma (\varphi - \vec{A}\cdot\vec{\beta}) 
                 + \frac{1}{8\pi}\iiint d^3\Omega~  (|\vec{E}|^2 - |\vec{H}|^2)
                   \Bigr]  .
\end{equation}
In this equation $d\tau$ and $d^3\Omega$ are the proper time and three-dimensional volume elements; $m$ and $e$ the mass and charge of the particle; $\vec\beta=\vec{v}/c$ its velocity and $\gamma= 1/\sqrt{1-\beta^2}$; while
$\varphi$ and $\vec{A}$ are the scalar and vector potentials, and $\vec{E}$ and $\vec{H}$ the electric and magnetic fields of the external electromagnetic field.  Using this action, and a few additional postulates, it is possible to derive all of classical electrodynamics. Similarly, in combination with other additional postulates, Eq.\ $\eqref{int:1}$ is used as an explicit or implicit input  to derive the basic equations of quantum mechanics and field theory.   However, in both cases, there are conceptual difficulties, such as ambiguities and infinities, which are still unsolved, despite the enormous practical success of present-day theory.

In his doctoral dissertation of 1919 Cornelius Lanczos showed that Eq.\ $\eqref{int:1}$ could in fact be \emph{derived} form a pure field theory, in which charged particles correspond to singularities in the Maxwell field interpreted as a biquaternion-analytic field generalizing to four complex dimensions the well-known Cauchy-Riemann theory of complex-analytic functions.$^{(1,2)}$  Consequently, instead of $\eqref{int:1}$ the most fundamental equation in \emph{Lanczos's electrodynamics} is the action integral 
\begin{equation} \label{int:2}
   \REA \frac{1}{8\pi} \iiiint d^3\Omega\,d\tau~ \CON{B}B ,
\end{equation}
where $B=\vec{E}+i\vec{H}$ is the total electromagnetic field of all particles and external fields, and the invariant scalar $\CON{B}B$ the squared modulus of this total field.\footnote{Since Lanczos's electrodynamics is a biquaternion (i.e., complex quaternion) field theory, we will use the quaternion formalism in this paper.  However, as we will not make any detailed calculations here, readers unfamiliar with quaternions (i.e., compounds such as $Q=s+\vec{V}$) can interpret the quaternion algebra as an explicit whole symbol formalism combining scalars and ordinary vectors, in which all entities are either 4-vectors, such as the 4-velocity $\mathcal{U}=\gamma(1-i\vec{\beta})$, the 4-potential $A=\varphi-i\vec{A}$, the 4-gradient $\nabla = \partial_{ict}+\vec{\nabla}$, and the hypersurface elements that will be used in 4-dimensional  integrations;  or 6-vectors such as the electromagnetic field $B=\vec{E}+i\vec{H}$.  The  quaternion conjugation operation, i.e., $\CON{Q}=\CON{s+\vec{V}}=s-\vec{V}$, is equivalent to the tensor operation of raising/lowering an index so that the product $\CON{Q}Q$ yields an invariant scalar.  A few more quaternion definitions will be recalled in footnotes.  For further details on quaternion notations and methods we refer the reader to the references given in Refs.\ 3 and 14.}

   In this paper it will be shown that not only is it possible to derive $\eqref{int:1}$ from Lanczos's action $\eqref{int:2}$, but that this derivation is unambiguous and devoid of infinities.  It will also be shown that in order to reach this conclusion it is necessary to properly define the boundary surface associated with the singularities, a field theoretical requirement which suggests a remarkable similarity between the structure of leptons and hadrons.\footnote{It should also be stressed that Lanczos's electrodynamics is the first example of a modern field theory in which there is no ``mass term'' in the fundamental Lagrange function, and where ``mass'' arises as a result of self-interaction or symmetry breaking.  See, e.g., Steven Weinberg, ``A model of leptons,'' \emph{Phys. Rev. Lett.} {\bf 19}, 1264 (1967).}

   In this respect this paper is therefore a continuation and a conclusion of the commentary on Lanczos's dissertation that we had written in 1994,\footnote{In that commentary we did not write Lanczos's action $\eqref{int:2}$ using the operator ``$\REA$'' but used a slightly more general formulation which is not needed here.} and which was published in the Lanczos collection,$^{(3)}$ together with a facsimile of Lanczos's handwritten dissertation.$^{(1)}$  Since then we have made considerable progress in our understanding of Lanczos's electrodynamics and its relation to standard classical and quantum electrodynamics.  In particular, one of the major problems we had in 1994 was that the sign of the mass coming out of Lanczos derivation, Eq.\ (10) in Ref.\ 3, page 2-21, was difficult to understand and to accept.  In fact that sign was correct, and our present understanding is that its interpretation --- which we give in this paper --- could be a major breakthrough in the explanation of the origin of infinities in electrodynamics.

\vspace{2\baselineskip}
{\bf \noindent 2. LANCZOS'S DERIVATION}
%--------------------------------------
\vspace{1\baselineskip}
\label{lan:0}

The most striking difference between Lanczos's action integral $\eqref{int:2}$ and the usual one $\eqref{int:1}$ is the absence of an explicit ``mass'' (or ``self-interaction'') term of the form $mc^2\int d\tau$, as well as the absence of an ``interaction'' term featuring the scalar product $e\Scal[\CON{A_e} \mathcal{U}]$ of the  4-potential $A_e$ of the external field by the 4-current $e\mathcal{U}$ of the particle:\footnote{The operator $\Scal\Q$ means that we take the scalar part of the bracketed quaternion expression.}  There is only a ``field'' term which has the same form as the third term in $\eqref{int:1}$. This is because Lanczos' electrodynamics is a pure field theory, which is fundamentally based on Maxwell's homogeneous equations
\begin{equation} \label{lan:1}
 \nabla B = 0  ,
\end{equation}
in contradistinction to the usual theory which is based on Maxwell's \underline{in}homogen\-eous equations where charges and currents are postulated to be the causal sources of the fields.\footnote{As Maxwell's equations written in the form $\eqref{lan:1}$ provide a generalization of the Cauchy-Riemann regularity conditions from complex numbers to biquaternions, Eq.\ $\eqref{lan:1}$ is the basis of powerful developments in hyper-complex analysis which, however, will not be needed in this paper.  For a recent advance in these developments, and references to earlier steps, see Ref.\ 4.}  Therefore, in Lanczos electrodynamics, there are neither charges or currents, but simply singularities which are changing their positions in three-dimensional space in any continuous manner.\footnote{However, while such motions correspond to standard classical electrodynamics, in which $\varphi$ and $\vec{A}$ are real, nothing prevents to consider worldlines in which singularities move into complex spacetime, a possibility that was investigated by Lanczos in his dissertation already,$^{(2,3)}$ and which can be shown to yield hadronic fields and interactions.$^{(5)}$}

For instance, writing the field of some particle as $B_i$, the action corresponding to its interaction with a given external field $B_e$ will be
\begin{equation} \label{lan:2}
 \REA \frac{1}{8\pi} \iiiint  d^3\Omega\,d\tau~ 
                         (\CON{B_i} + \CON{B_e})
                         (     B_i  +      B_e )
\end{equation}
which trivially leads to the expression 
\begin{equation} \label{lan:3}
 \REA \frac{1}{8\pi} \iiiint d^3\Omega\,d\tau~
                    \Scal\Bigl[\CON{B_i} B_i 
                           + 2 \CON{B_e} B_i 
                           +   \CON{B_e} B_e 
                         \Bigr]  .
\end{equation}
Therefore, in Lanczos's electrodynamics, provided all integrals are feasible and finite, the first and second terms of this expression should yield the mass and interaction terms of the usual action integral $\eqref{int:1}$, which in quaternion notation translates to 
\begin{equation} \label{lan:4}
S=  -  mc^2 \int    d\tau~
    -  e    \int    d\tau~  \Scal\Bigl[\CON{A_e} \mathcal{U}\Bigr]
    + \REA \frac{1}{8\pi}  \iiiint d^3\Omega\,d\tau~ \CON{B_e} B_e .
\end{equation}

\vspace{2\baselineskip}
{\bf \noindent 3. THE USUAL INFINITE ELECTROMAGNETIC SELF-INTERACTION}
%---------------------------------------------------------------------
\vspace{1\baselineskip}

\label{vol:0}

In practice, if one takes for $A_i$ and $B_i$ the Li\'enard-Wiechert potential and field of an arbitrarily moving particle (which in Lanczos's electrodynamics are interpreted as the potential and field associated with a moving point singularity), i.e.,\footnote{$A_i$ and $B_i$ are the potential and field at the space-time point $\mathcal{X}$ produced by a charge $e$ located at the point $\mathcal{Z}$. The 4-velocity $\mathcal{U} = \dot{\mathcal{Z}}$, as well as the retarded distance $\xi = \Scal[\CON{\mathcal{U}}(\mathcal{X} - \mathcal{Z})]$, are evaluated at the retarded proper time $\tau_r$. The binary operator $\vect$ means that after making the quaternion product the scalar part is discarded so that the result is a vector.}
\begin{equation} \label{vol:1}
      A_i = e \frac{\mathcal{U}}{\xi} ~~~ ~~~ , ~~~ ~~~
      B_i = \CON\nabla \vect A_i ,
\end{equation}
and if the integrations in Eq.\ $\eqref{lan:3}$ are made in the ``standard way,'' that is as volume integrals over the whole of three-space, one finds out that in the general case both the mass and the interaction terms diverge.  This is readily seen by calculating the mass term in the rest-frame of the particle, i.e.,
\begin{equation} \label{vol:2}
   \REA \frac{1}{8\pi} \iiiint d^3\Omega\,d\tau~  \CON{B_i} B_i
   = \int d\tau~ e^2 \int_{\xi_1 \rightarrow 0}
                         ^{\xi_2 \rightarrow \infty}d\xi~\frac{1}{2\xi^2} ,
\end{equation}
which, referring to $\eqref{int:1}$ or $\eqref{lan:4}$, gives for the mass the divergent expression
\begin{equation} \label{vol:3}
   mc^2 = - e^2 \int_{\xi_1 \rightarrow 0}
                    ^{\xi_2 \rightarrow \infty}d\xi~\frac{1}{2\xi^2}
        =   e^2 (\lim_{\xi_2 \rightarrow \infty}\frac{1}{2\xi_2}
               - \lim_{\xi_1 \rightarrow 0}\frac{1}{2\xi_1}) = - \infty .
\end{equation}
According to the usual interpretation, which is prevalent since the end of the 19th century, this expression has two defects for our purpose:  Not only does it lead to an infinite electromagnetic mass when $\xi_1 \rightarrow 0$, but its \emph{sign} is wrong.   However, if for some reason (as is arbitrarily done in quantum electrodynamics) the infinite term is discarded, one obtains a mass which is finite and of the correct sign, provided $\xi_2$ is kept finite instead of made infinite.  This observation (which we make with the benefit of hindsight) leads to the further observation that the infinite term in $\eqref{vol:3}$ could not be there in the first place.  Indeed, if the integral $\eqref{vol:2}$ is made using distribution theory (or, equivalently in the present case, Hadamard's theory of ``finite parts of an integral''$^{(8)}$) the result is mathematically equal to what is obtained by discarding the infinite term.  This can be seen by making a change of parametrization such that the angular integration is made over half instead of the full sphere.  The 3-space integration in $\eqref{vol:2}$ gives then
\begin{equation} \label{vol:4}
   \REA \frac{1}{8\pi} \iiint d^3\Omega~  \CON{B_i} B_i
   =   e^2 ~\mathsf{H}
              \hspace{-4.35mm}\int_{\xi_2 \rightarrow -\infty}
                         ^{\xi_2 \rightarrow +\infty}d\xi~\frac{1}{4\xi^2}
          = - \lim_{\xi_2 \rightarrow \infty}\frac{e^2}{2\xi_2}  ,
\end{equation}
where the last step comes from taking Hadamard's finite part, see Ref.\ 8, page 787.

\vspace{2\baselineskip}
{\bf \noindent 4. CALCULATION OF THE SELF-INTERACTION TERM}
%----------------------------------------------------------
\vspace{1\baselineskip}
\label{sel:0}

While the previous section's calculation of a finite value for the mass term is fully satisfactory from a mathematical point of view, a physically more intuitive reason for that finite result derives from the pure field theoretical character of Lanczos's electrodynamics. The reason is that the ``standard way'' of calculating action integrals as volume integrals does not take the full nature of electromagnetic singularities into account --- a point that Lanczos strongly emphasized in his doctoral dissertation: \emph{The four-dimensional integrations should be made in the spirit of field theory, that is as hypersurface integrals.}

   In principle, that is always possible since the homogeneous Maxwell's equations $\eqref{lan:1}$ enable to use Gauss's theorem to transform volume integrals into surface integrals.  Moreover, calculating four-dimensional integrals as was done in Sec.~3, i.e.,  by going to the rest frame without taking causality explicitly into account, may lead to incorrect results  --- something that is less likely to occur when calculating surface integrals which by necessity require the boundary conditions to be explicitly considered and taken into account.

   The difficulty with this approach is that defining and calculating four-dimensio\-nal hypersurface integrals can be conceptually and technically difficult.  In fact, in his doctoral dissertation, Lanczos was not able (or possibly did not even attempt) to perform such integrations in the general case and to show that all three terms in $\eqref{lan:3}$ could be finite and in agreement with $\eqref{int:1}$.  Neither did he later return to this problem.

In our case, we were fortunate to find out that Paul Weiss (the particularly brilliant first Ph.D.\ student of Dirac) rediscovered the importance of general hypersurfaces in the calculation of four-dimensional quantum action integrals,$^{(6)}$ a point that opened the way to the later theories of Tomonoga, Schwinger, \emph{et al.}, which led to modern quantum electrodynamics --- see references in Ref.\ 3.  In the same vein, Paul Weiss also developed powerful methods for the explicit calculation of four-dimensional surface integrals, using for this purpose the biquaternion algebra to make explicitly the spinor decomposition of four-vectors and six-vectors.$^{(7)}$  This formalism is particularly appropriate for the present problem because Lanczos's electrodynamics is a biquaternion field theory, and because Weiss's methods are designed to deal with arbitrarily accelerated motions.  However, possibly at the expense of some technical difficulties, the same calculations may also be done using other methods or formalisms, which is why we do not show the details but only the major steps in the calculations.

We therefore return to Eq.\ $\eqref{lan:3}$ and rewrite it using Maxwell's equations $\eqref{lan:1}$, the relation between the four-potential and the field $\eqref{vol:1}$, as well as Gauss's theorem to transform the first two terms into hypersurface integrals, i.e.,
\begin{equation} \label{sel:1}
 \REA \frac{1}{8\pi} 
                    \Scal\Bigl[ \iiint  \CON{A_i}\, d^3\Sigma\, B_i
                           + 2  \iiint  \CON{A_e}\, d^3\Sigma\, B_i 
                           +   \iiiint  d^3\Omega\,d\tau~ \CON{B_e} B_e
                         \Bigr]  .
\end{equation}

   The next step is to chose an appropriate hypersurface enclosing the world line of the particle between two points corresponding to the proper times $\tau_1$ and $\tau_2$.  For this purpose, although due to Gauss's theorem any hypersurface bounding a causally connected four-volume could be used in principle, the most convenient one is Weiss's proper tube of constant retarded radius $\xi_2$,\footnote{Throughout this paper we use $\xi_2$ for the retarded radius of a tube or sphere which is finite or such that $\xi_2 \rightarrow \infty$, while we use $\xi_1$ for a radius such that $\xi_1 \rightarrow 0$.} closed at both ends by the light cones erected at the times $\tau_1$ and $\tau_2$.  The advantages of this surface is that all calculations can be done exactly and consistently, because everything is expressed in terms of the invariant variables $\xi$ and $\tau$ --- the retarded distance and proper time which also appear in $\eqref{vol:1}$ --- and because the closure of both ends of the proper tube by light cones insures that the boundaries define a causally connected subspace of spacetime.\footnote{Before Weiss the proper tube was used by Bhabha.$^{(9)}$  However, neither of them did close its ends with light cones because --- following Dirac$^{(10)}$ --- their intent was to let the radius of the tube tend to zero at the end of the calculation.}

Calculating the first (i.e., mass or self-interaction) term in $\eqref{sel:1}$, the contribution from the proper tube is found to be
\begin{equation} \label{sel:2}
 \REA \frac{1}{8\pi} 
                    \iint\limits_{sphere}\int_{\tau_1}^{\tau_2}
                    \Scal \Bigl[ \CON{A_i}\, d^3\Sigma_{tube}\, B_i \Bigr]
                 = -\frac{e^2}{2\xi_2} \int_{\tau_1}^{\tau_2} d\tau  ,
\end{equation}
and that from the two light cones
\begin{equation} \label{sel:3}
                       \REA \frac{1}{8\pi} 
                       \iint\limits_{sphere}\int_{\xi_1}^{\xi_2}
                       \Scal \Bigl[ \CON{A_i}\, d^3\Sigma_{cones}\, B_i \Bigr]
                 = e^2 \ln (\frac{\xi_2}{\xi_1}) {\bigg|}_{\tau_2}
                 - e^2 \ln (\frac{\xi_2}{\xi_1}) {\bigg|}_{\tau_1}
                      \equiv 0 .
\end{equation}
Therefore, by making a proper hypersurface integration, we find that while the contribution of the tube is finite (as could be expected), the divergence which was $1/\xi_1$ for $\xi_1 \rightarrow 0$ in $\eqref{vol:3}$ where the ``standard method'' was used, is now logarithmic and such that it cancels out for any value of $\xi_1$ and $\xi_2$.  Therefore, the contribution from the two end cones, Eq.\ $\eqref{sel:3}$, is also \emph{zero} in the limit $\xi_1 \rightarrow 0$, so that the total value of the mass term is equal to $\eqref{sel:2}$, which is \emph{finite}.

Consequently, we have reached the conclusion that in Lanczos's electrodynamics the mass term is exactly equal to
\begin{equation} \label{sel:4}
   \REA \frac{1}{8\pi} \iiiint d^3\Omega\,d\tau~  \CON{B_i} B_i
             = - \frac{e^2}{2\xi_2}  \int_{\tau_1}^{\tau_2} d\tau ,
\end{equation}
where $\xi_2$ is the radius of a tube of constant retarded distance surrounding the worldline.  Hence, as already observed in Sec.~3, if $\xi_2 \not\rightarrow \infty$  we may identify the factor ${e^2}/{2\xi_2}$ with the non-zero mass $m$ in $\eqref{int:1}$, a procedure which requires a justification that will be given after calculating the interaction term.

\vspace{2\baselineskip}
{\bf \noindent 5. CALCULATION OF THE INTERACTION TERM}
%-----------------------------------------------------
\vspace{1\baselineskip}
\label{act:0}

To calculate the second (i.e., interaction) term in $\eqref{sel:1}$ it is necessary to remember that our goal is to derive the action integral $\eqref{lan:4}$ from Lanczos's action $\eqref{lan:2}$, and that the usual action integral, i.e., Eqs.\ $\eqref{lan:4}$ or $\eqref{int:1}$, implicitly assumes that the external field is ``given,'' that is non-affected by the motion of the particle, and influencing the motion of the particle solely by its value at the position of the particle.  Consequently, if the particle is assumed to be of vanishingly small size, there is an implicit assumption that the external field is slowly varying in the region close to it.

   In principle this condition is satisfied by assuming (for the sake of the derivation) that the external field is constant, which has the advantage to simplify the calculation.  However, if we postulate that Lanczos's electrodynamics is the more fundamental theory from which the usual action is derived, it is better to assume that the external field may vary in space and time, albeit in such a way that its variation over the region of integration close to the worldline is negligible.  Since in our case this region is defined by the proper tube and cones this leads to the conditions\footnote{We write ``$Q \ll R$'' to imply that the components of the quaternions $Q$ and $R$ satisfy a condition such that ``$|Q_n| \ll |R_n|$.''}
\begin{equation} \label{act:1}
    \xi_2 \frac{\partial}{\partial \xi} A_e   \ll A_e 
        ~~~ \Big| ~~~ \forall \tau \in [ \tau_1,\tau_2 ] ,
\end{equation}
and
\begin{equation} \label{act:2}
    \xi \frac{\partial}{\partial \tau} A_e   \ll A_e 
        ~~~ \Big| ~~~ \forall \xi \in [ \xi_1,\xi_2 ]  .
\end{equation}

The contribution from the proper tube is then found to be
\begin{equation} \label{act:3}
                 \REA \frac{1}{4\pi} 
                 \iint\limits_{sphere}\int_{\tau_1}^{\tau_2}
                 \Scal \Bigl[ \CON{A_e}\, d^3\Sigma_{tube}\, B_i \Bigr]
            = -e \int_{\tau_1}^{\tau_1} d\tau~
      \Scal \Bigl[ \CON{A_e} (\mathcal{U} + i\xi_2\dot{\mathcal{U}}) \Bigr] ,
\end{equation}
and that from the two light cones
\begin{equation} \label{act:4}
                 \REA \frac{1}{4\pi} 
                 \iint\limits_{sphere}\int_{\xi_1}^{\xi_2}
                     \Scal \Bigl[ \CON{A_e}\, d^3\Sigma_{cones}\, B_i \Bigr]
                 = e (\xi_2 - \xi_1) \Scal \Bigl[\CON{A_e}\,\mathcal{U} \Bigr]
                       ~ {\bigg|}_{\tau_1}^{\tau_2} .
\end{equation}

   As can be seen, both contributions are finite if $\xi_2 \not\rightarrow \infty$.  Moreover, there is no divergence when letting $\xi_1 \rightarrow 0$.  On the other hand, there is an additional contribution of the form $i\xi_2\dot{\mathcal{U}}$ in $\eqref{act:3}$, i.e., an ``acceleration correction'' to the four-velocity $\mathcal{U}$, which is absent in the usual action $\eqref{lan:4}$.  As a matter of fact, this extra contribution gave us a lot of trouble in the 1994 version of our commentary on Lanczos's dissertation.$^{(3)}$  However, in that commentary, we made the mistake of not closing the proper tube with end cones.  Indeed, if we integrate by part the $\CON{A_e}\dot{\mathcal{U}}$ term in $\eqref{act:3}$, and then add $\eqref{act:4}$ to $\eqref{act:3}$, we get for the total interaction term
\begin{equation} \label{act:5}
 \REA \frac{1}{4\pi} 
                 \iiint \Scal \Bigl[ \CON{A_e}\, d^3\Sigma_{total}\, B_i \Bigr]
                    = -e \int_{\tau_1}^{\tau_1} d\tau~
      \Scal \Bigl[ (\CON{A_e} + i \xi_2 \dot{\CON{A_e}}) \mathcal{U} \Bigr]
\end{equation}
where the acceleration correction has disappeared and, instead, a contribution of the form $i\xi_2\dot{\CON{A_e}}$ has been added to the external potential.

   If we now compare this final result to the interaction term in the usual action $\eqref{lan:4}$ we see that this new contribution can be neglected because of the assumption $\eqref{act:2}$, provided the radial integration is restricted to the interval $\xi \in [0,\xi_2]$ and $\xi_2 \not\rightarrow \infty$.  Consequently, taking these assumptions into account, we see that in Lanczos's electrodynamics the interaction term is
\begin{equation} \label{act:6}
 \REA \frac{1}{4\pi} 
                 \iiiint d^3\Omega\,d\tau~ 
              \Scal \Bigl[ \CON{B_e}\, B_i \Bigr]
                    = -e \int_{\tau_1}^{\tau_1} d\tau~
              \Scal \Bigl[ \CON{A_e} \mathcal{U} \Bigr] ,
\end{equation}
i.e., equal to the corresponding term in the usual action integral of classical electrodynamics.

\vspace{2\baselineskip}
{\bf \noindent 6. THE PROPER BOUNDARY POSTULATE}
%-----------------------------------------------
\vspace{1\baselineskip}
\label{pro:0}

All along this paper we have used the adjective ``proper'' to qualify algebraic quantities such as the proper time, and geometrical entities such as the proper tube, as a means to specify that these objects are consistently defined in accord with the principles of relativity and causality.

On the other hand, we have not yet given much consideration to the physical interpretation of the domains of integration and to their boundaries in relation to Lanczos's action principle $\eqref{int:2}$, or more specifically $\eqref{lan:3}$ in our derivation of the action integral $\eqref{int:1}$.  In particular, we have implicitly assumed that the integrations in $\eqref{lan:3}$ should be made over all of three-space, i.e., that $\xi_2 \rightarrow \infty$, while the identification of $\eqref{sel:4}$ and $\eqref{act:6}$ with the mass and interaction terms in $\eqref{int:1}$ strongly suggests that the $\xi$-integration should be truncated at a finite value of $\xi_2$.  Therefore, the initial assumption that all integrations should be made over all of three-space should be questioned, because otherwise the mass term $\eqref{sel:4}$ will be zero, and the interaction term $\eqref{act:5}$ possibly infinite.

  In fact, quoting from Lanczos's dissertation: ``If the field-theoretical point of view is correct, the boundaries must also have a field-theoretical meaning,'' see Ref.\ 1, Chap.\ 8. This implies that the hypersurface surrounding the worldline that was used in the previous sections should have such a meaning.  For practical reasons, and for consistency with the principles of relativity and causality, we took a proper tube closed at both ends by light cones. Nevertheless, due to Maxwell's homogeneous equations and Gauss's theorem, this hypersurface is equivalent to any other causally connected closed surface surrounding the worldline:  The only difference is that after making the integrations some numerical factors in $\eqref{sel:4}$ or $\eqref{act:6}$ could possibly be different.  We therefore formulate the following postulate:

\begin{quote}
{\bf Postulate:}  \emph{(i) The proper boundary hypersurface to be used for each term containing a singular field $B_i$ in the action integral $\eqref{lan:3}$, i.e.,
\begin{equation} \label{lan:3-bis}
 \REA \frac{1}{8\pi} \iiiint d^3\Omega\,d\tau~
                    \Scal\Bigl[\CON{B_i} B_i 
                           + 2 \CON{B_e} B_i 
                           +   \CON{B_e} B_e 
                         \Bigr]  ,
\end{equation}
is a proper tube of constant retarded distance closed at both ends by light cones; and (ii) the integration of these terms should be made over the inside of the subspace bounded by this tube.}
\end{quote}

   This postulate provides a geometrical picture of a charged particle as a singularity enclosed in a surface: the proper two-sphere of constant retarded distance which transported along the worldline defines the proper tube.  Consequently, such a particle has a basic property: An ``inside'' and an ``outside'' which can be related to the concept of \emph{coupling}.  Indeed, the second clause in the postulate, which implies that the $\xi$-integrals in $\eqref{sel:4}$ and $\eqref{act:5}$ have to be made between 0 and $\xi_2$, is equivalent to the assumption that the external field $B_e$ couples exclusively with that part of the field $B_i$ which is inside the proper sphere, just like it is only that part of $B_i$ which couples with itself in the self-interaction term.\footnote{This is why we have mnemonically written $A_i$ and  $B_i$ for the Li\'enard-Wiechert potential and field of the moving singularity.}

   However, this geometrical picture should not be taken as a model of the electron.  In particular, the proper sphere is not the border of some kind of a physical object but an abstract boundary, which can be seen by noting that there is no discontinuity in the potential or the field  at this boundary.  Moreover, there is no relation between this picture and the Abraham-Lorentz electron modeled as a localized distribution of charge.  On the other hand, if we let the radius $\xi_2 \rightarrow 0$, the expressions $\eqref{sel:4}$ and $\eqref{act:5}$ for the mass and interaction terms tend towards their usual expressions calculated in standard electrodynamics for a point charge, including the obnoxious infinite electromagnetic mass.\footnote{But, if we let $\xi_2 \rightarrow \infty$ we ultimately enclose all singularities in the universe: there is no truly external field and all masses are zero, so that one cannot derive the usual action integral anymore.}

\vspace{2\baselineskip}
{\bf \noindent 7. INTERPRETATION OF THE MASS OF A SINGULARITY}
%-------------------------------------------------------------
\vspace{1\baselineskip}
\label{mas:0}

In Secs.~4 to 6 we have seen that Lanczos's electrodynamics applied to the motion of a singularity in an external field unambiguously leads to the expressions $\eqref{sel:4}$ and $\eqref{act:6}$ for the mass and interaction terms, which fully agree with the corresponding terms in the usual action integral of classical electrodynamics $\eqref{lan:4}$, provided we make for the ``mass'' $m$ the assignment
\begin{equation} \label{mas:1}
                     mc^2  =  \frac{e^2}{2\xi_2}   ,
\end{equation}
which implies that $\xi_2$ is equal to half the ``classical electron radius,''
\begin{equation} \label{mas:2}
                     r_e  =  \frac{e^2}{mc^2}  ,
\end{equation}
if $m$ is taken as the mass of an electron or positron.

    This leads to the question of how to interpret this assignment, because (as seen in Sect.\ 3) the usual interpretation of the mass of a charged particle as its ``electromagnetic mass,'' i.e., the mass associated with the energy in the electromagnetic field surrounding the particle, is incompatible with Lanczos's derivation: It would lead to a negative mass, which, besides, would be infinite if the energy density of the electromagnetic field is integrated between the location of the singularity and infinity.

In fact, it is only if the integration of the self-interaction term is made under the constraint that we have a true biquaternion-analytic singularity, i.e., such that we have to take Hadamard's finite part, or else to replace the volume integral by a hypersurface integral, that we get a mass of the correct sign; and only if the integration is made over a region bounded by a finite radius $\xi_2$ that we get a non-zero value for the mass $\eqref{mas:1}$.

Therefore, when calculating the self-interaction term, the proper sphere of radius $\xi_2$ surrounding the singularity at every moment in its motion along the world line is acting as a boundary such that the energy within the proper sphere is equal to $-mc^2$, which because of the minus sign in $\eqref{int:1}$ yields the mass given in $\eqref{mas:1}$, while the energy in the field outside the proper sphere is equal to $+mc^2$.

For this reason, the well-known fact that the mass given by expression $\eqref{mas:1}$ corresponds to the energy in the electromagnetic field surrounding the particle integrated between $\Oh r_e$ and infinity is to be regarded as fortuitous, even though this occurrence may have some deep significance since the total energy obtained by integrating  over the whole space is zero.

On the other hand, the negative energy within the proper sphere can be seen as a kind of ``electromagnetic mass defect,'' which may be interpreted as a ``binding energy'' explaining why singularities are possible and stable in Lanczos electrodynamics.\footnote{This interpretation would also explain why new phenomena, which are not described by the action integral $\eqref{int:1}$ or $\eqref{lan:4}$, are possible for interaction energies larger than $mc^2$.}  However, considering that Lanczos's electrodynamics is a field theory engaging the reconsideration of many fundamental concepts related to electrodynamics and elementary particles,\footnote{In this paper we have assumed that the singularity is a simple Li\'enard-Wiechert pole.  Nothing prevents to consider more complicated singularities,$^{(5)}$ or clusters of several singularities at a distance on the order of $r_e$.} it is better at this stage not to take such interpretations literally.

In this spirit, the interpretation of the mass $m$ given by $\eqref{mas:1}$ is that it is simply the \emph{inertial mass} of the singularity, i.e., the mass appearing in $\eqref{int:1}$ if that action integral is taken as the fundamental equation of classical electrodynamics.\footnote{The mass $m$ is then the factor multiplying the acceleration in the equation of motion which derives from the Lagrange function associated with the action integral $\eqref{int:1}$.}

   This interpretation is confirmed by other applications of Lanczos's electrodynamics, for instance the derivation of the Abraham-Lorentz-Dirac equation of motion.$^{(10)}$  Since this equation includes radiation reaction it cannot be derived from the action integral $\eqref{int:1}$ without further assumptions.  On the other hand, it is straightforward to derive it starting from Lanczos's electrodynamics.$^{(11)}$  In the course of this derivation it is found, as in the present case, that all integrals are finite and that there is no direct relation between $m$ and the usual concepts of ``electromagnetic'' and ``mechanical'' mass.\footnote{In general the terms ``mechanical mass,'' ``material mass,'' and ``inertial mass'' are interchangeable and equal to the ``experimental mass.''  However, when dealing with the concept of renormalization, as will be done below, the mechanical mass is used to refer to a non-electromagnetic contribution to the experimentally measured mass.}  Moreover, it is found that the radius $\xi_2$ to be used to get the inertial mass through an assignment of the form $\eqref{mas:1}$ is not $\tfrac{1}{2} r_e$ but $\tfrac{2}{3} r_e$, which means that the radius appearing in $\eqref{mas:1}$ is not a ``fundamental length,'' but a length on the order of the classical electron radius whose precise value depends on the problem under consideration.

  In other words, the assignment $\eqref{mas:1}$ has to be understood as a simple normalization step by which the quantity obtained by integrating the self-interaction term in Lanczos's action integral $\eqref{lan:3}$ is set equal to the ``experimental'' mass, i.e., the \emph{inertial mass} in the sense of D'Alembert, see Ref.\ 12, Chap.\ IV.

    Therefore, contrary to the practice initiated in classical electrodynamics by Dirac,$^{(10)}$ and in quantum electrodynamics by Kramers,$^{(13)}$ there is no need for \underline{re}normali\-zation in Lanczos's electrodynamics.  For instance, in the present derivation of the mass appearing in the usual action integral $\eqref{int:1}$, there is no need for the standard rule due to Dirac and Kramers, i.e.,
\begin{equation} \label{mas:3}
                     m_{exp}  =  m_{mec} + m_{ele}  ,
\end{equation}
which comes from the iterative process of starting some classical or quantum calculation by using (either explicitly or implicitly) a principle such as the action integral $\eqref{int:1}$ --- in which $m$ is interpreted as a ``bare'' or ``mechanical'' mass $m_{mec}$ to be corrected (i.e., \underline{re}normalized)  by a (possibly infinite) contribution $m_{ele}$ at the end of the process.  This is because our derivation is  based on Lanczos's action integral $\eqref{int:2}$ --- where there is no ``mass'' term --- or equivalently\footnote{See the Appendix of Lanczos's dissertation and the Preface in Ref.\ 2.} Maxwell's homogeneous equations $\eqref{lan:1}$ --- where there is no ``source'' term --- so that there is neither a mass nor a charge to \underline{re}normalize.

Finally, a truly unambiguous interpretation of mass in Lanczos's electrodynamics is provided by Weiss's derivation of the Abraham-Lorentz-Dirac equation of motion.$^{(7)}$  Indeed, using the quaternion methods he had developed for that purpose, Weiss obtained a fully independent (as well as mathematically and physically rigorous) derivation of that equation which avoids several pitfalls of Dirac's derivation  --- something that is still not appreciated today.$^{(14)}$  Moreover, since Weiss used only Maxwell's homogeneous equations and made only surface integrations, his derivation is in full accord with Lanczos's electrodynamics, of which he was totally unaware. In particular, he introduced a polygenic work function (i.e., a non-integrable differential, see Ref.\ 12, Chap.\ I) which enabled him to obtain the Abraham-Lorentz-Dirac equation of motion by means of a variational principle directly related to D'Alembert's principle, so that the mass appearing in his derivation is necessarily the inertial mass.\footnote{In Weiss's original derivation the mass is infinite. However, by closing the proper tube and consistently dealing with the singular terms the mass turns out to be finite.$^{(14)}$}

\newpage

\vspace{2\baselineskip}
{\bf \noindent 8. CONCLUSION}
%----------------------------
\vspace{1\baselineskip}
\label{con:0}

   In this paper we have successfully completed the proof sketched by Lanczos, in his dissertation of 1919, of his claim that the usual action integral of classical electrodynamics $\eqref{int:1}$ can be derived from his more fundamental action integral $\eqref{int:2}$.

In the course of this derivation we have found that the mass term is exact and finite, and that the derivation of the interaction term requires the supplementary conditions $\eqref{act:1}$  and $\eqref{act:2}$.  These conditions can be united in the four-dimensional requirement 
\begin{equation} \label{con:1}
    r_e \frac{\partial}{\partial x_n} A_e   \ll A_e 
        ~~~ \Big| ~~~ \forall \tau \in [ \tau_1,\tau_2 ] , 
\end{equation}
which means that the external field must be slowly varying with respect to all four coordinates $x_n \in \{ \tau, x_1,x_2,x_3 \}$ over the full extent of the proper tube whose radius is on the order of the classical electron radius $r_e$.  Thus, as this requirement is equivalent to the well-known conditions for the internal consistency of classical electrodynamics, we have succeeded in driving both the usual action integral and these conditions from Lanczos's electrodynamics.\footnote{We recall that it is often considered that actually, because of quantum effects, classical electrodynamics is already not applicable for fields at least $\alpha^{-1} \approx 137$ times smaller than implied by $\eqref{con:1}$.  Our opinion is that if Lanczos's electrodynamics is taken as a fundamental theory which encompasses classical electrodynamics, as well as some aspects of quantum theory and general relativity (see Ref.\ 5), its consequences must be consistent on their own.} %  In this perspective, the quantum effects must be considered as \emph{corrections} on the order of $\alpha \approx 1/137$ to Lanczos's theory.}

  In this derivation an essential role is played by Lanczos's identification of Maxwell's homogeneous equations with a four-dimensional generalization of the Cauchy-Riemann analyticity conditions, which led him to postulate that electrodynamics is a pure field theory analytic over the biquaternion algebra.  This postulate, summarized by Eq.\ $\eqref{lan:1}$, enables to replace four-volume integrals by three-surface integrals and to handle singularities in a consistent manner.  It then turns out that the self-interaction integral leads to a finite mass term.%,\footnote{The way in which the finite mass is obtained is reminiscent of the calculation of the flux of the electric field through a closed surface, which leads to a finite charge without problems, whereas the volume integral over the enclosed charge density leads to an incorrect result unless the singularities at the origin of the field are properly taken care of.} and that all calculations can be done in an explicit and exact manner using the methods developed by Paul Weiss to make the integrations.

   Consequently, in Lanczos's field-theoretical approach to electrodynamics the boundary conditions and the correct choice of the domains of integration are of fundamental importance, and essentially equivalent to the definition of the elementary physical objects described by the theory.  This implies that the proper tube of finite radius surrounding the worldline has such a meaning. Since this tube is obtained by transporting a proper sphere along the worldline one is led to a remarkable observation, namely that the singularities in Lanczos's electrodynamics are necessarily associated in a fundamental way to a proper sphere, and that for electrons this proper sphere has a radius on the order of $r_e = 2.817 \times 10^{-15}$~m, which is (as has often been noticed, e.g., Ref.\ 15, page 791) on the same order as the electromagnetic radius of protons and other elementary particles.\footnote{If this proper sphere is assimilated to a ``bag,'' which is readily observable for hadrons, the reason why it is not observable for electrons is that electronic bags are occupied by a single singularity, while there are two or more singularities in hadronic bags so that the electromagnetic radii of hadrons are non-zero. For an application of the proper sphere concept to the mass spectrum of quarks and leptons see Ref.\ 16.}

\vspace{2\baselineskip}
{\bf \noindent REFERENCES} 
% ------------------------
\vspace{1\baselineskip}
\begin{enumerate}
%
% Set spacing for FoF reference list
%\setlength{\parskip}{-1.4mm}

\item Kornel Lanczos, \emph{Die Funktionentheoretischen Beziehungen der Max\-well\-schen  Aethergleichungen --- Ein Beitrag zur  Relativit\"ats- und Elektronentheorie} (Verlagsbuchhandlung Josef N\'emeth, Budapest, 1919). Reprinted in \emph{Cornelius Lanczos Collected Published Papers With Commentaries},  W.R.\ Davis \emph{et al.}, eds.\ (North Carolina State University, Raleigh, 1998) Vol.\ {\bf VI}, pages A-1 to A-82. Web site \underline{http://www.physics.ncsu.edu/lanczos}.

\item Cornelius Lanczos, \emph{The relations of the homogeneous Maxwell's equations to the theory of functions --- A contribution to the theory of relativity and electrons} (1919, Typeseted by Jean-Pierre Hurni with a preface by Andre Gsponer, 2004); e-print arXiv:physics/0408079 available at\\ \underline{http://arXiv.org/abs/physics/0408079}.

\item A. Gsponer and J.-P. Hurni, ``Lanczos's functional theory of electrodynamics --- A commentary on Lanczos's PhD dissertation, in \emph{Cornelius Lanczos Collected Published Papers With Commentaries},  W.R.\ Davis \emph{et al.}, eds.\ (North Carolina State University, Raleigh, 1998) Vol. {\bf I}, pages 2-15 to 2-23; e-print arXiv:math-ph/0402012 available at\\ \underline{http://arXiv.org/abs/math-ph/0402012}.

\item K. Sano, ``Another type of Cauchy's integral formula in complex Clifford analysis,'' \emph{Tokyo J. Math.} {\bf 20}, 187 (1997).

\item A. Gsponer, ``On the physical interpretation of singularities in Lanczos-Newman electrodynamics;'' e-print arXiv:gr-qc/0405046 available at\\ \underline{http://arXiv.org/abs/gr-qc/0405046}.

\item P. Weiss, \emph{Proc. Roy. Soc. A} {\bf 156}, 192 (1936);  {\bf 169}, 102 (1938); {\bf 169}, 119 (1938).

\item P. Weiss, ``On some applications of quaternions to restricted relativity and classical radiation theory,'' \emph{Proc.  Roy.  Irish.  Acad.} {\bf 46}, 129 (1941).

\item R. Courant and D. Hilbert, \emph{Methods of Mathematical Physics} (Interscience Publ., New York, 1962)  Vol. {\bf 2}.

\item H.J. Bhabha, ``Classical theory of mesons,'' \emph{Proc. Roy. Soc. A} {\bf 172}, 384 (1939).

\item P.A.M. Dirac, ``Classical theory of radiating electrons,'' \emph{Proc. Roy. Soc. A}  {\bf 167}, 148 (1938).

\item A. Gsponer, ``Straightforward derivation of the Abraham-Lorentz-Dirac equation of motion,'' Report ISRI-03-11 (2003).

\item C. Lanczos, \emph{The Variational Principles of Mechanics} (Dover, New York, 1949, Fourth edition, 1970).

\item H.A. Kramers, ``Die Wechselwirkung zwischen geladenen Teilchen und Strahlungsfeld,'' \emph{Nuovo Cim.} {\bf 15}, 108 (1938).

\item A. Gsponer and J.-P. Hurni, ``Paul Weiss's derivation of the Abraham-Lorentz-Dirac equation of motion,'' Report ISRI-04-08 (2004).

\item J.D. Jackson, \emph{Classical Electrodynamics} (Wiley, New York, Second edition, 1975).

\item A. Gsponer, J.-P. Hurni, ``Non-linear field theory for lepton and quark masses,'' Hadronic Journal {\bf 19}, 367 (1996); e-print arXiv:hep-ph/0201193 available at \underline{http://arXiv.org/abs/hep-ph/0201193}.

\end{enumerate}

\end{document}